\documentclass[11pt]{article}
\usepackage{amsmath}
\usepackage{amsfonts}
\usepackage{amssymb}
\usepackage{graphicx}
\nonstopmode

\begin{document}
\title{\bf The driven oscillator}
\author{T.B. Smith
\\
\ {\scriptsize Department of Physical Sciences}\\{\scriptsize The Open University, Walton Hall, Milton Keynes, MK7 6AA, UK}}
\maketitle
\begin{abstract}
We consider the quantum mechanics of an harmonic oscillator when it is driven either by an external random (white noise) force or when its frequency is sinusoidally time-dependent, either varying slowly (adiabatically) or at about twice the natural frequency (parametrically). We focus on finding transition probabilities and expectations, for which it proves convenient to utilize the Wigner-Weyl formulation associating an operator $\hat{A}$ with its phase space transform, $\hat{A}\longleftrightarrow A(p,q)$ and in particular the density matrix $\hat{\rho}$ with its corresponding Wigner function $\rho_w$ defined by $\hat{\rho}(t)\longleftrightarrow h \rho_w(p,q;t)\,.$  It is shown that for long times, whatever the oscillator's initial state, a white noise driving force randomizes over time the distribution with respect to phase for any operator $\hat{\Phi}\longleftrightarrow \Phi(\phi)$ and in particular for the phase operator $\hat{\phi}\longleftrightarrow \phi$ itself, where $\phi$ is the polar angle in the phase plane. We also consider, in the long-time limit, the randomizing effect of the white noise force for expectations of $\hat{\phi}^2$ and of any operator $\hat{\Omega}\longleftrightarrow \Omega (R)$, where $R$ is the radial coordinate in the phase plane, the partner to $\phi$ such that $\frac{p^2}{2m} + \frac{m \omega^2}{2}q^2 \equiv \frac{\hbar \omega}{2} R^2$. When there is no external force acting, but there is a time-dependent frequency $\omega^2(t)=\omega_0^2(1 + \epsilon (t))$, where $\epsilon$ is small, then, if $\epsilon$ varies slowly, transition probabilities are unaffected, but for a parametric oscillator $\epsilon$ varies sinusoidally at a frequency of about twice $\omega_0$ and these probabilities are strongly time-dependent.
\end{abstract}


\section{Introduction}
Analyzing the quantum harmonic oscillator---or more generally any polynomial Hamiltonian quadratic in $\hat{p}$ and $\hat{q}$ with time-dependent coefficients---is about as old as quantum mechanics itself. A recent broad approach \cite{lima} summarizes and develops an analysis based on dynamical invariants \cite{lewis} to calculate the time evolution of wave functions. One can also approach the problem using path summation \cite{khandekar}.  A third alternative (at least when the frequency and/or mass depend on time)
is solution by ansatz \cite{ciftja}. These methods can be, and have been, used to calculate the effects on its wave function of a given time-dependence in an oscillator's Hamiltonian.  In the present paper, however, the problem is transformed to phase space using the Wigner-Weyl association. It appears that this method is particularly suited to calculating time-dependent expectations and transition probabilities, and that is our focus here.

There are of course models of the quantum mechanical response of a small system such as an oscillator to the action of a heat bath \cite{Hondaetc} or also with with external forces acting \cite{PGetc}. Here we analyze exactly the response of certain operators associated with an oscillator, to the randomizing effects of a white noise random external force, for which the transformation to the Wigner-Weyl picture proves useful.

In general, the association of an operator, $\hat{A} \longleftrightarrow A(p,q)$ with a function on the phase plane can take many forms \cite{cohen}, but the commonest, and in many ways the simplest, is that of Wigner and of Weyl \cite{weyla,weylb,wigner}. This paper utilizes {\em only} the Wigner-Weyl correspondence. The density matrix $\hat{\rho}$ and phase operator $\hat{\phi}$ (see reference \cite{dhsbook}) are well-defined quantum mechanical objects with the correspondences $\rho_w(p,q;t) \longleftrightarrow \hat{\rho}(t) /h$ and $\phi(p,q){ \longleftrightarrow} \hat{\phi}$, where $\rho_w(p,q;t)$ is the time-dependent Wigner function and $\phi$ is effectively the polar angle in the phase plane $(p,q)$.  Here I mostly consider the time dependence of operators related to the one-dimensional harmonic oscillator under the influence of a time-dependent force or when its frequency is time-dependent. The general Hamiltonian for these cases is
\begin{equation}\label{ham1}
  \hat{H}(t) \equiv \frac{\hat{p}^2}{2m} + \frac{1}{2}m \omega^2 (t)\hat{q}^2 - \lambda(t) \hat{q} \,.
\end{equation}
A great advantage of using the Wigner-Weyl phase space formulation for the oscillator is that when the quantum mechanics is `unwrapped' the dynamics is purely classical. A price to pay, however, is that to make the final result properly quantum mechanical one must perform one or more integrals over phase space.

Section \ref{WignerWeyl} reviews briefly the basic association $\hat{A}\longleftrightarrow A(p,q)$.  The time-dependence of $A$ and its application to the harmonic oscillator, equation (\ref{ham1}), is discussed section \ref{time}. The material in sections \ref{WignerWeyl} and \ref{time} is not new. In section \ref{noiseforcing} we consider the action of a white noise Gaussian random force $\lambda(t)$ whose ensemble average vanishes but with $\overline{ \lambda(t_1)\lambda(t_2)} = \mu\, \delta(t_1 - t_2)$.  We use an analysis of classical Brownian motion in phase space \cite{smith79} to find the ensemble averaged quantum mechanical propagator for the Wigner function. From this one can calculate transition probabilities for the oscillator. In particular we calculate explicitly the probability that the oscillator leaves the ground state after a time $t$ under the action of $\lambda(t)$.

Section \ref{phase} is devoted in the first instance to the time evolution of the phase angle operator $\hat{\phi}$ and then, more generally, to operators $\hat{\Phi} \longleftrightarrow \Phi(\phi)$ under the influence of the stochastic force.  An expression is derived for the expectation ${\rm Tr}( \hat{\rho}(t)\hat{\Phi})$ when the initial state is the oscillator's ground state $|h_0\rangle\langle h_0|$.  More generally it is also shown that, whatever the initial state may be, at long times, the distribution of $\Phi(\phi)$ is randomized with respect to the angle $\phi$.
Section \ref{phase} also considers the long time effects (for any initial state $\hat{\rho}(0)$) of the random force on the expectation ${\rm Tr}(\hat{\rho}(t)\hat{ \Omega})$ of operators $\hat{\Omega} \longleftrightarrow \Omega (R)$ where $R$ is defined by the Weyl transform of the free oscillator's Hamiltonian such that $p^2/(2m) + m \omega^2 q^2/2 = \hbar \omega R^2/2$. Section \ref{phase} ends by considering, in the long time limit, the influence of the random force on the behaviour of the expectation ${\rm Tr} (\hat{\rho}(t) \hat {\phi}^2)$, for any initial state $\hat{\rho}(0)$.  Although $\hat{\phi}^2$ itself has a Weyl transform depending on {\em both} $\phi$ and $R$, this expectation becomes randomized in time with respect to $\phi$.

Section \ref{tdepfreq} considers, approximately, two special cases of a time-dependent frequency $\omega(t)$, but with no other external influence. On the one hand when $\omega(t)$ is slowly varying (the adiabatic limit) a simple time dependent phase shift results, generally of limited physical consequence. However, for a parametric oscillator, the frequency varies sinusoidally at about twice the oscillator's natural frequency. In that case it is shown that when the oscillator is initially in the ground state the probability of it staying there decreases, in the limit, exponentially in time.

Section \ref{disc} gives a brief discussion.
\section{The Wigner-Weyl picture}\label{WignerWeyl}
There are many possible formulations of quantum mechanics in phase space \cite{cohen}.
Generally, they can be related \cite{smith06,lee} to that of Wigner and Weyl,
for which we shall adopt a formal but efficient notation \cite{degroot}.
Denoting the Weyl transform of an operator $\hat{A}$ by $A(p,q)$, or sometimes by $(\hat{A})(p,q)$, it is given by
\begin{equation}\label{weyl1}
  \hat{A} \longleftrightarrow {\rm Tr}(\hat{A}\,\hat{\Delta}(p,q))\equiv A(p,q)\equiv (\hat{A})(p,q)
\end{equation}
where
\begin{eqnarray}\label{weyl}
\mathrm{}    \hat{\Delta}(p,q) & = &
    \int_{-\infty}^{\infty}\frac{ {\rm d} p' {\rm d}q'}{h}
    \,{\rm e}^{-\frac{\rm i}{\hbar} (p'q - q'p )}
    \,\hat{D}(p',q') \nonumber\\
    \\
    & = & \int_{-\infty}^{\infty}
    {\rm d}x \,\,{\rm e}^{\frac{\rm i}{\hbar}p\, x}\,\,
   |q + \frac{x}{2} \rangle \langle q - \frac{x}{2}|  \nonumber
\end{eqnarray}
and $\hat{D}$ is the Weyl operator (\cite{klauder}),
\begin{equation}\label{Dpq}
    \hat{D}(p,q) = {\rm e}^{\frac{\rm i}{\hbar}(p\, \hat{q} - q
\,\hat{p})}\,.
\end{equation}
Formally $\hat{\triangle}$ has the the properties \cite{degroot} that its trace
is unity, that
\begin{equation}\label{weyl3}
    \int_{-\infty}^{\infty}\frac{{\rm d}p\,{\rm d}q}{h}
    \hat{\Delta}(p,q) = 1\,,
\end{equation}
and that
\begin{equation}\label{weyl4}
    {\rm Tr} \left( \hat{\Delta}(p,q) \hat{\Delta}(p',q') \right)
    = h \delta(p - p') \delta(q - q')\,.
\end{equation}
Using this, equation (\ref{weyl1}) can be inverted to give
\begin{equation}\label{weylone}
    \hat{A} = \int_{-\infty}^\infty \frac{{\rm d}p\,{\rm d}q}{h}
    A(p,q) \hat{\Delta}(p,q) \quad\mbox{and}\quad {\rm Tr}\hat{A}=\int_{-\infty}^\infty \frac{{\rm d}p\,{\rm d}q}{h}
    A(p,q)\,.
\end{equation}
From these properties one can also show that, for two operators
$\hat{A}$ and $\hat{B}$,
\begin{equation}\label{weyl5}
    {\rm Tr}(\hat{A}\hat{B}) = \int_{-\infty}^{\infty}\frac{{\rm d}p\, {\rm d}q}{h}
    A(p,q) B(p,q)\,,
\end{equation}
and that the Weyl transform of the product $\hat{A} \hat{B}$ is
\begin{equation}\label{weyl6}
 \hat{A} \hat{B}  \longleftrightarrow {\rm Tr}\left(\hat{A} \hat{B} \hat{\Delta}(p,q) \right)=
    A(p,q)\, {\rm exp}\left[ \frac{{\rm i}\hbar}{2}
    \left( \frac{\partial^*}{\partial q} \frac{\partial}{\partial p}\, -\,
     \frac{\partial^*}{\partial p} \frac{\partial}{\partial q} \right) \right] B(p,q)\,,
\end{equation}
where the starred operators act to the left on $A(p,q)$.

The  Wigner-Weyl association can be expressed in terms of either $\hat{D}$ or $\hat{\Delta}$, whichever is
more convenient, for they are related by the first of equations (\ref{weyl}), or equivalently, by
\begin{equation}\label{dee}
  \hat{D}(p,q) = \int_{-\infty}^{\infty}\frac{ {\rm d} p' {\rm d}q'}{h}
    \,{\rm e}^{-\frac{\rm i}{\hbar} (p'q - q'p )} \,\hat{\Delta}(p',q')\,.
\end{equation}
Algebraically, the relation between $\hat{D}$ and $\hat{\Delta}$ can be expressed \cite{royer,smith06} as
\begin{equation}\label{ddelta}
  \hat{\Delta}(p,q) = 2\, \hat{D}(2 p,2 q)\, \hat{\Pi},\,\,\, \text{so that}\,\,\,
  \hat{D}(p,q) = \frac{1}{2}\, \hat{\Delta}(p/2,q/2)\, \hat{\Pi},
\end{equation}
where $\hat{\Pi}$ is the parity operator.

The Weyl transform makes the fundamental association
\[\hat{D}(p',q') = {\rm e}^{\frac{\rm i}{\hbar}(p'\, \hat{q} - q'
\,\hat{p})} { \longleftrightarrow}\, {\rm e}^{\frac{\rm i}{\hbar}(p'\,q - q'\,p)} \,, \]
so that, sensibly, $\hat{1} { \longleftrightarrow} 1$, $F(\hat{p}){ \longleftrightarrow} F(p)$ and $F(\hat{q})  {\longleftrightarrow} F(q)$. But functions that mix $\hat{p}$ and $\hat{q}$ are more complicated. For instance $\hat{p}\, \hat{q}{ \longleftrightarrow} p\,q -{\rm i}\hbar/2 $ and $\hat{q} \hat{p}{ \longleftrightarrow} p\,q +{\rm i}\hbar/2 $ so that the Weyl operator corresponding to $p \, q$ is $(\hat{p}\, \hat{q}+ \hat{q}\, \hat{p})/2 $. The Weyl transform of the Hamiltonian operator, equation (\ref{ham1}), takes the classical form
\begin{equation}\label{ham2}
  \hat{H}(t)\longleftrightarrow  \frac{p^2}{2m}+\frac{1}{2}m \omega^2 (t)q^2 - \lambda(t) q
\end{equation}
Finally, the the Wigner function is defined as
\begin{equation}\label{wigt}
  \rho_w(p,q;t)\equiv \frac{1}{h} {\rm Tr}\big(\hat{\rho}(t) \hat{\Delta}(p,q) \big)
\longleftrightarrow \frac{1}{h}\hat{\rho}(t)\,,
\end{equation}
so that by (\ref{weyl5}),
\begin{equation}\label{wignert}
  {\rm Tr} \big(\hat{\rho}(t) \hat{A}\big) = \int_{-\infty}^{\infty} {\rm d}p \, {\rm d}q \, \rho_w(p,q;t) A(p,q)\,.
\end{equation}
\section{Time development}\label{time}
The evolution of the Wigner function under the action of a time-dependent Hamiltonian is is well-known \cite{smith06,degroot,smith78}. Wave functions evolve according to
\begin{equation}\label{wftime}
  |\psi_t \rangle = \hat{U}_t | \psi\rangle,
\end{equation}
where the unitary time evolution operator ${\hat U}_t$ is governed by the equation
\begin{equation}\label{evo}
  {\rm i} \hbar \frac{\partial}{\partial t} \hat{U}_t =
  \hat{H}_t \hat{U}_t\,,
\end{equation}
and we allow for explicit time-dependence in the Hamiltonian $\hat{H}_t$.
Then the time-dependent density matrix $\hat{\rho(t)}$ is given by
\begin{equation}\label{opev}
 \hat{\rho}(t) = {\hat U}_t \, \hat{\rho}(0)\,{\hat U}^{\dagger}_t\,,
\end{equation}
and its Weyl transform is
\begin{equation}\label{opevo}
    \rho_w(p,q;t) = \int {\rm d}p' {\rm d}q' P_w(p,q,t|p',q',0)\rho_w(p',q';0)
\end{equation}
where $P_w(|)$ is the Wigner propagator defined by
\begin{equation}\label{wignerprop}
    P_w(p,q,t|p',q',0) = \frac{1}{h}{\rm Tr}(\hat{U}_t^\dag \hat{\Delta}(p,q)\hat{U}_t \hat{\Delta}(p',q')).
\end{equation}
In particular, it is easy to show from the properties in section \ref{WignerWeyl} that
\begin{equation}\label{property1}
    \int{\rm d}p' {\rm d}q' P_w(p,q,t|p',q',0)=\int{\rm d}p\, {\rm d}q P_w(p,q,t|p',q',0)=1
\end{equation}
and
\begin{equation}\label{property2}
    P_w(p,q,0|p',q',0) = \delta(p-p')\delta(q-q')\,.
\end{equation}
From definition (\ref{wignerprop}) and equation (\ref{evo}) we can differentiate this propagator with respect to time to get its equation of motion as the Weyl transform of a commutator, as follows:
\begin{equation}\label{commutator}
  {\rm i}\hbar \frac{\partial}{\partial t}P_w(p',q',t|p,q,0)=
\frac{1}{h}
 \left \{\, [\hat{H}_t,{\hat U}_t \hat{\Delta}(p,q) {\hat U}_t^\dag]\,\right \}(p',q')
\end{equation}
 The Weyl transform of ${\hat H}_t$ given by equation (\ref{ham2}) is
quadratic in $p$ and $q$ and so has no derivatives with respect to $p$ and/or $q$ of order higher than second.  Applying equation (\ref{weyl6}) to (\ref{commutator}) is straightforward: collecting terms gives
\begin{eqnarray}\label{propmotion}
    \lefteqn{\frac{\partial}{\partial t}P_w(p,q,t|p',q',0) = }\nonumber\\ & = & \frac{\partial}{\partial q}H_t(p,q)\frac{\partial}{\partial p}P_w(p,q,t|p',q',0)-\frac{\partial}{\partial p}H_t(p,q)
    \frac{\partial}{\partial q}P_w(p,q,t|p',q',0)\,.
\end{eqnarray}
This describes classical motion under the action of $H_t(p,q)$. The solution that satisfies condition (\ref{property2}) is
\begin{equation}\label{classicalP}
  P_w(p,q,t|p',q',0) = \delta(p - p(t|p',q',0))\delta(q-q(t|p',q',0))\,,
\end{equation}
where $(p(t|p',q',0),q(t|p',q',0))$ is the classical phase space solution for momentum and position under the action of Hamiltonian $H_t(p,q)$ such that $\big(p(0|p',q',0),q(0|p',q',0)\big)= (p',q')$.  This solution also obeys equation (\ref{property1}), by direct integration in the first instance and, in the second, by recognizing that the Jacobian $\partial(p,q)/ \partial(p',q')$ is unity when $(p,q)$ and $(p',q')$ are related by the classical motion implied by equation (\ref{classicalP}).  For the cases we are considering, the equations for classical motion are
\begin{equation}\label{motion}
  p(t) = m \dot{q}(t) \quad \mbox{and}\quad \dot{p}(t) =
- m \omega^2 q(t) - \lambda\,.
\end{equation}
where $\omega$ and/or $\lambda$ may depend on time.
\section{Forcing by stationary white noise}\label{noiseforcing}
\subsection{Time dependent external force}
As shown in section \ref{time}, the time evolution for the harmonic oscillator can be found by solving the classical equations of motion. For instance, when the oscillator is subjected to a time-dependent force only, then $\lambda$ depends on time and $\omega$ and $m$ are constants. Expressed in terms of dimensionless variables $(x,y) = (p/(\hbar \alpha),\alpha q)$, where $\alpha ^2 \equiv m \omega /\hbar $, equations (\ref{motion}) can be written
\begin{equation}\label{xymotion}
  \frac{\rm d}{{\rm d} t}(x+ {\rm i}y) - {\rm i} \omega (x + {\rm i}y) =
  - \frac{1}{\alpha \hbar} \lambda(t) \,,
\end{equation}
Writing $x(t)$ and $y(t)$ for $ x(t|x_0,y_0,0)$ and  $y(t|x_0,y_0,0)$, the solutions with initial values $x_0$ and $y_0$, we have
\begin{equation}\label{xysolution}
  x(t) + {\rm i}\,y(t) = {\rm e}^{{\rm i} \omega t}\left[(x_0 + {\rm i}y_0)
  - \frac{1}{\alpha \hbar}\int_{0}^{t}{\rm d}s \, {\rm e}^{-{\rm i} \omega s}\,
  \lambda(s)\right]\,.
\end{equation}
For example, when \[\lambda(t) = \lambda_0 \sin(\omega t + \theta)\,, \]
\begin{equation*}\label{xyt}
  x(t)+\rm{i}y(t)=R_0  {\rm e}^{{\rm i}(\omega t + \theta_0)}
  +\frac{{\rm i} \lambda_0}{2 \alpha \hbar \omega}\Big( \omega t\,{\rm e}^{{\rm i}(\omega t + \theta)} - \sin(\omega t){\rm e}^{- {\rm i}\theta} \Big)\,,
\end{equation*}
where
\[x_0 + {\rm i} y_0 \equiv R_0 {\rm e} ^{{\rm i} \theta_0}. \]
\subsection{The motion}
We model the effects of a stationary random force by supposing the Hamiltonian has the form
\begin{equation}\label{randomham}
  \hat{H} = \frac{\hat{p}^2}{2 m}+\frac{m \omega^2}{2}\hat{q}^2 - \lambda(t)\, \hat{q}\,,
\end{equation}
where $\lambda(t)$ is a white noise stationary Gaussian process \cite{wax} with the particular ensemble averages,
\begin{equation}\label{noise}
  \overline{\lambda(t)} = 0 \quad\mbox{and}\quad \overline{ \lambda(t_1)\lambda(t_2)} = \mu\, \delta(t_1 - t_2)\,.
\end{equation}
One can, of course, use equation (\ref{xysolution}) and take the relevant ensemble averages, but this has already been done for the classical theory of Brownian motion
\cite{wax}. In that theory, a traditional approach is through the Langevin equations. In one dimension these are
\begin{equation}\label{brownian}
  p= m \dot{q}\quad \mbox{and}\quad \dot{p} = F(p,q) - \beta p + \lambda(t)\,,
\end{equation}
where $\lambda(t)$ is the random force and $F$ may depend on $p$ or $q$. The friction term, $- \beta p$ represents the effect of a heat bath and cannot be modelled by any Hermitian single particle Hamiltonian. Here we shall be considering the limit $\beta \rightarrow 0$.
Denoting the the ensemble average of the classical propagator for Brownian motion by an overline, from equation (\ref{classicalP}) we have
\[\overline{ P_w{(p,q,t|p_0,q_0,0)}}=\overline { \delta(p-p(t|p_0,q_0,0)) \delta(q-q(t|p_0,q_0,0))} \equiv W(p,q,t|p_0,q_0,0)  \]
where $W(\cdot|\cdot)$ is the conditional probability density for $(p,q)$ at time $t$ given the initial conditions $(p_0,q_0)$ at $t=0$.

Classical Brownian motion is random, stationary and Markovian, and it is characterized by the conditional probability density $W(p,q,t|p_0,q_0,0)$. Reference \cite{smith79} gives the following efficient expression that is generally approximate, but \emph{exact} when $F(p,q)$ is at most linear in $p$ and/or $q$:
\begin{eqnarray}\label{brownianone}
    \lefteqn{W(p,q,t|p_0,q_0,0) \simeq  \int \frac{{\rm d}a{\rm d}b}{(2 \pi)^2}\,
  {\rm exp}\big({\rm i}a(q-q(t)\big){\rm exp}\big({\rm i}b(p-p(t)\big)}\nonumber\\
   && \hspace{5cm} \times \,\, {\rm exp}\left[-\frac{\mu}{2}  \int_0^t {\rm d}s
  \Big(\partial_{p_s}\big(aq(t)+bp(t)\big)\Big)^2\right]\hspace{4cm}
\end{eqnarray}
where $\mu$ characterizes the random force $\lambda(t)$, $(p(t),q(t))$ is shorthand for the
solutions \\$\big(p(t| p_0,q_0,0),q(t| p_0,q_0,0)\big)$ to (\ref{brownian}) (most generally with $\beta \neq 0$), and $\partial_{p_s}$ is the partial derivative with respect to the momentum $p_s$ at the intermediate time $0 \leq s\leq t$.  Thus $\mu$ occurs explicitly in equation (\ref{brownianone}) and $\beta$ occurs implicitly through the solution $(p(t),q(t))$. If the classical Brownian particle were to thermalize after long times \cite{smith79} then the relation  $\mu/2=\beta k T/m $ must obtain, where $k$ is Boltzmann's constant and $T$ the temperature. Here we choose to consider the limit $\beta \rightarrow 0$, so modelling an oscillator driven only by the random force $\lambda(t)$, equations (\ref{randomham}) and (\ref{noise}). Another way to look at this model (see reference \cite{PGetc}) is via the WFPE (Wigner-Fokker-Planck equation) for an oscillator in thermal equilibrium with a thermal heat bath in the limit of high temperatures but small friction such that the product $\beta k T$ remains finite.

In our picture we require a solution for the unforced harmonic oscillator, $p(t)$ and $q(t)$, with vanishing $\beta$. These are, for $0\leq s \leq t$,
\begin{equation}\label{clone}
  p(t)=p_s \cos\left(\omega(t-s)\right)-m \omega q_s \sin\left(\omega(t-s) \right)
\end{equation}
and
\begin{equation}\label{cltwo}
  q(t)= \frac{p_s}{m \omega} \sin(\omega(t-s))+q_s \cos(\omega(t-s))\,.
\end{equation}
Using this information to evaluate (\ref{brownianone}) and replacing $\simeq$ by $=$, for this result is exact, gives
\begin{eqnarray}\label{osc}
  \lefteqn{W(p,q,t|p_0,q_0,0) =  \int \frac{{\rm d}a{\rm d}b}{(2 \pi)^2}\,
  {\rm exp}\big({\rm i}a(q-q(t)\big)}\nonumber\\
   &&\hspace{1cm}\times \,{\rm exp}\big( {\rm i}b(p-p(t) \big) {\rm exp}\left[-\frac{\mu}{2}\int_0^t {\rm d}s \left( \frac{a}{m \omega} \sin\omega s+ b \cos\omega s \right)^2 \right],
\end{eqnarray}
wherein $p(t)$ and $q(t)$ refer to the solutions with initial conditions $p_0,q_0$.
The simplest case is the free particle, for which $\omega\rightarrow 0$. Taking that limit in (\ref{osc}) gives, in detail,
\begin{eqnarray}\label{free}
 \hspace{3cm} \lefteqn{\hspace{-4cm} W(p,q,t|p_0,q_0,0) =}\nonumber\\ && \hspace{-3cm} \int \frac{{\rm d}a{\rm d}b}{(2 \pi)^2}\,
  {\rm exp}\left({\rm i}a\big(q-q_0-\frac{p_0}{m}t\big)\right){\rm exp}\big({\rm i}b(p-p_0)\big)
    \, {\rm exp}\left[-\frac{\mu}{2}\int_0^t {\rm d}s \left(\frac{a}{m}s +b\right)^2 \right]\,.\hspace{0cm}
\end{eqnarray}
All the integrals in (\ref{osc}) and (\ref{free}) can be easily evaluated if necessary.
\subsection{Transition probabilities}\label{transprob}
The probability for transition between states $|\psi_1\rangle$ and $|\psi_2\rangle$ is
\begin{eqnarray}\label{trans}
  \hspace{-1cm}|\langle \psi_2 |\hat{U}_t |\psi_1 \rangle|^2 &=& { \rm Tr}\left(|\psi_2\rangle\langle\psi_2|\hat{U}_t|\psi_1\rangle\langle \psi_1|\hat{U}_t^\dag \right)\nonumber\\
  &=& \int \frac{{\rm d}p\,{\rm d}q}{h}\int {\rm d}p'{\rm d}q'\, \big(|\psi_2\rangle\langle\psi_2|\big)(p,q)P_w (p,q,t|p',q',0)\big(|\psi_1\rangle\langle\psi_1|\big)(p',q')\,,
\end{eqnarray}
where $P_w$ is the Wigner propagator, (\ref{wignerprop}). In particular, for the random driving force (\ref{noise}), the ensemble averaged propagator is
\begin{equation*}
  W (p,q,t|p',q',0)=\overline{P_w(p,q,t|p',q',0)}\,.
\end{equation*}
The corresponding ensemble averaged transition probability is
\begin{equation}\label{average}
 \overline{ |\langle \psi_2 |\hat{U}_t |\psi_1 \rangle|^2} = \int \frac{{\rm d}p\,{\rm d}q}{h}\int {\rm d}p'{\rm d}q'\, \big(|\psi_2\rangle\langle\psi_2|\big)(p,q)W (p,q,t|p',q',0)\big(|\psi_1\rangle\langle\psi_1|\big)(p',q')
\end{equation}
where, for an oscillator, $W$ is given by (\ref{osc}). Writing $(\partial_x,\partial_y)$  for partial derivatives with respect to $x$ and $y$, where $(x,y)= (p/(\hbar \alpha),\alpha q)$, and $\alpha^2 = m \omega/\hbar$, that expression for $W$ can be formally rewritten as
\begin{equation}\label{Wone}
  W(p,q,t|p',q',t)= {\rm exp}\left[\frac{N}{2}\int_0^{\omega t}{\rm d} \theta\, (\sin \theta\, \partial_y + \cos\theta\, \partial_x)^2\right]\delta(p-p(t))\delta(q-q(t))\,,
\end{equation}
with
\begin{equation}\label{defs}
  \delta(p-p(t))\delta(q-q(t))=\frac{1}{\hbar}\,\delta(x-x(t)) \delta(y-y(t))\quad\mbox{and the definition}\quad N \equiv \frac{\mu}{m \omega^2 \hbar}\,.
\end{equation}
In (\ref{defs}) $(p(t),q(t))$ is the solution for the free oscillator with initial conditions $(p',q')$, that is to say $p(t)=p(t|p',q',0)$ and $q(t)=q(t|p',q',0)$, equations (\ref{clone}) and (\ref{cltwo}), so this classical propagator conserves the oscillator's energy.

As an example let us suppose that the oscillator is initially in the ground state, $|h_0\rangle$, and seek the probability in time that it stays there so that $|\psi_1\rangle=|\psi_2\rangle=|h_0\rangle$, where
\begin{equation}\label{groundstate}
  \langle \xi|h_0\rangle= (\alpha/ \sqrt{\pi})^{1/2}{\rm exp}\left(- \alpha^2 \xi^2/2\right)\quad\mbox{and}\quad  \alpha^2= \frac{m \omega}{\hbar}\,.
\end{equation}
The Weyl transform of the initial (and final) state is easily found to be
\begin{equation}\label{ground}
  \left(|h_0 \rangle\langle h_0|\right)(p,q)=2 \,{\rm exp}\left(- (x^2 +y^2)\right)=2\, {\rm exp}(-R^2)\,.
\end{equation}
This sole dependence on $R$ suggests that it will be convenient sometimes to integrate with respect to plane polar coordinates $(R,\phi)$. This can be effected in the integrals by the equivalences
\[{\rm d}p\,{\rm d}q = \hbar\, {\rm d}x \,{\rm d}y=\hbar\, R\, {\rm d} R\, {\rm d} \phi\,, \]
and the free motion in these coordinates is
\[\delta(p-p(t))\delta(q-q(t))=\frac{1}{\hbar R}\delta(R-R')\delta(\phi-\phi'-\omega t)\,. \]
Therefore, performing the integrals over $R'$ and $\phi'$, gives
\begin{eqnarray}\label{utee}
  \overline{|\langle h_0 |\hat{U}_t |h_0 \rangle|^2 } &=& \int \frac{{\rm d}p\,{\rm d}q}{h}\int {\rm d}p'{\rm d}q'\, \big(|h_0\rangle\langle h_0|\big)(p,q)W (p,q,t|p',q',0)\big(|h_0\rangle\langle h_0|\big)(p',q')\nonumber \\
   &=&   4 \int \frac{{\rm d}p\,{\rm d}q}{h}\,{\rm exp}(- R^2)\, {\rm exp}\left[\frac{N}{2}\int_0^{\omega t}{\rm d} \theta\, (\sin \theta\, \partial_y + \cos\theta\, \partial_x)^2\right]\,{\rm exp}(- R^2)\,.
\end{eqnarray}
We can now make the Fourier decomposition
 \begin{equation}\label{fourier}
   {\rm exp}(-R^2) = \int\frac{{\rm d}k_1 {\rm d}k_2}{4 \pi}\,{\rm exp}{\left(-\frac{k_1^2 + k_2^2}{4} \right)}{\rm exp}\big({\rm i}(k_1 x + k_2 y)\big)
 \end{equation}
so leading to the expression
\begin{multline}\label{quadone}
  \overline{|\langle h_0 |\hat{U}_t |h_0 \rangle|^2 } =4\int \frac{{\rm d}x\,{\rm d}y}{2 \pi}{\rm exp}(-R^2)\int\frac{{\rm d}k_1 {\rm d}k_2}{4 \pi}\,{\rm exp}{\left(-\frac{k_1^2 + k_2^2}{4}\right)}{\rm exp}\big({\rm i}(k_1 x + k_2 y)\big) \\ \times {\rm exp}\left[-\frac{N}{2}\int_0^{\omega t}{\rm d} \theta\, (k_1 \cos \theta +k_2 \sin \theta )^2\right]\,. \hspace{3cm}
\end{multline}
Expression (\ref{quadone}) is a four-fold quadratic integral and thus can be evaluated exactly. We can start with the double integral with respect to $k_1$ and $k_2$. The standard Gaussian form for this is
\begin{equation}\label{quadtwo}
  \int_{-\infty}^\infty {\rm d}x^n {\rm exp}\left[- \frac{1}{2}\, \mathbf{x}^T \cdot \mathbf{A} \cdot\mathbf{x}\right] {\rm exp}[{\rm i}\, \mathbf{J}^T \cdot \mathbf{x}]=\sqrt{\frac{(2\pi)^n}{{\rm det} \mathbf{A}}}\,\,{\rm exp} \left[- \frac{1}{2}\, \mathbf{J}^T \cdot \mathbf{A}^{-1} \cdot\mathbf{J} \right]
\end{equation}
where $\mathbf{A}$ is a real symmetric $n$ by $n$ matrix. This sort of integral finds application, for example, in evaluating path sums in field theory \cite {kaku}, often in the limit  of infinite $n$ In the present case, with $n=2$, $\mathbf{x}^T= (k_1,k_2)$, and $\mathbf{J}^T= (x,y)$, we have
\begin{equation}\label{matrixone} \mathbf{A}=\left(
    \begin{array}{cc}
      \frac{1}{2}+N S_{11}& \frac{N}{2} S_{12} \\ \\
      \frac{N}{2}S_{12} & \frac{1}{2}+N S_{22} \\
    \end{array}
  \right)\quad\mbox{where}\quad \mathbf{S} =\left(
                                  \begin{array}{cc}
                                    \int_0^{\omega t}{\rm d}\theta\cos^2\theta & \int_0^{\omega t}{\rm d}\theta \sin 2\theta \\ \\
                                    \int_0^{\omega t}{\rm d}\theta \sin 2\theta & \int_0^{\omega t}{\rm d}\theta\sin^2\theta  \\
                                  \end{array}
                                \right)\,.
\end{equation}
The inverse of $\mathbf{A}$ is
\begin{equation}\label{matrixtwo}
\mathbf{A}^{-1}=\frac{1}{{\rm det} \mathbf{A}} \left(
  \begin{array}{cc}
    \frac{1}{2}+N S_{22}& -\frac{N}{2} S_{12} \\ \\
      -\frac{N}{2}S_{12} & \frac{1}{2}+N S_{11} \\
  \end{array}
\right).
\end{equation}
Combining (\ref{quadone}) and (\ref{quadtwo}) then gives
\begin{equation}\label{quadthree}
  \overline{|\langle h_0 |\hat{U}_t |h_0 \rangle|^2 } =\frac{1}{\pi\sqrt{ {\rm det} \mathbf{A}}} \int {\rm d}x\,{\rm d}y{\rm \,exp}(-R^2) {\rm exp} \left[- \frac{1}{2}\, \mathbf{J}^T \cdot \mathbf{A}^{-1} \cdot\mathbf{J} \right]\,.
\end{equation}
This has the standard form
\begin{equation}\label{quadB}
  \int_{-\infty}^\infty {\rm d} x^n \, {\rm exp}\left[- \frac{1}{2}\, \mathbf{x}^T \cdot \mathbf{B} \cdot \mathbf{x}\right] =\sqrt{\frac{(2 \pi)^n}{{\rm det} \mathbf{B}}}\,,
\end{equation}
where, in this case, $n=2$ and $\mathbf{x }= (x,y)$. For (\ref{quadB}) it is easy to see that
\[ \mathbf{B} = \mathbf{A}^{-1} + 2 \mathbf {I}\, \]
where $\mathbf{I}$ is the unit matrix. Finally then,
\begin{eqnarray}\label{quadfour}
  \overline{|\langle h_0 |\hat{U}_t |h_0 \rangle|^2 } &=&\frac{1}{\pi\sqrt{ {\rm det} \mathbf{A}}}\,\frac{2 \pi}{\sqrt{ {\rm det}( 2 \mathbf{I} + \mathbf{A}^{-1}})}=\frac{2}{{\rm det}(\mathbf{I}+2 \mathbf{A})} \nonumber \\
  &=& \frac{1}{\sqrt{\left(1 + \frac{N \omega t}{2}\right)^2 -\left(\frac{N}{2}\right)^2 \sin^2 \omega t}}\,,
\end{eqnarray}
where $N$ is given by (\ref{defs}) and we have expanded the determinant using expressions (\ref{matrixone}).
\section{Phase}\label{phase}
\subsection{Definition}
Generally in the Wigner-Weyl picture, the time-dependent average of an operator $\hat{A}$ with respect to the state $\hat{\rho}$ is given by (\ref{wignert}) with (\ref{opevo}), where, for an oscillator, the evolution is classical, equation (\ref{classicalP}). In particular consider the the Weyl quantization $\hat{\phi}{\longleftrightarrow} \phi(p,q)$, of the harmonic oscillator phase in the plane $(p,q)$.  Properties of $\hat{\phi}$ have been considered previously \cite{dhsbook,sdh92,dhsrims1,lynchrev}. It is a bone-fide bounded self-adjoint operator on Hilbert Space. As befits an angle, its spectrum must be limited to a range of $2 \pi $, but it appears that a published proof \cite{dhsbook,dhsrims1} only exists limiting the spectrum to a range of $3 \pi $.  There may, however, be an unpublished proof \cite{Daniel}---using the `method of wedges' defined in section 9.4.5 of \cite{dhsbook}---limiting the spectrum to a range of $2 \pi$. In any case, numerical calculations representing $\phi $ by the finite matrix $\langle h_m | \hat{\phi} |h_n \rangle, \,\, 0 \leq (m,n) \leq N$ in terms of the harmonic oscillator energy eigenstates, indicate---as $N$ is increased up to several hundred---a smooth convergence from within to a uniform spread of eigenvalues over an interval of length $2 \pi$. In the following we consider the time-dependence of $\hat{\phi}$ when the oscillator is driven by a white noise stochastic force, equations (\ref{randomham}) and (\ref{noise}).

We choose to define $\hat{\phi}$ in terms of the harmonic oscillator creation operator
\[ \hat{a}^\dag =
\frac{1}{\sqrt{2}}\left(\alpha\hat{q} - {\rm i}\frac{\hat{p}}{\alpha \hbar}\right){ \longleftrightarrow} a^*(p,q)=\frac{1}{\sqrt{2}}\left(\alpha q - {\rm i}\frac{p}{\alpha
\hbar}\right) \]
where, for an oscillator having mass $m$, $\alpha^2 \equiv m \omega/\hbar \,\,.$
Then, in terms of the dimensionless variables $(x,y) = (\frac{p}{\hbar \alpha},\alpha q)$ the Weyl transform of $\hat{a}\dag$ is $\frac{-\rm i}{\sqrt{2}}R\, {\rm e}^{{\rm i} \phi}$ where $R^2 = x^2 + y^2$ and the the energy of a free oscillator is $\frac{\hat{p}^2}{2 m} + \frac{m
\omega^2}{2}\hat{q}^2 \stackrel{w}{ \longleftrightarrow} \frac{p^2}{2 m} + \frac{m
\omega^2}{2}q^2 = \frac{\hbar \omega}{2}R^2$. The Weyl transform of $\phi$ involves an integral over the phase plane $(p,q)$. So if we use plane polar coordinates $(R,\phi)$, where
${\rm d}p \,{\rm d}q/h = R {\rm d}R {\rm d}\phi / 2 \pi$, we must restrict $\phi$ to a range of $2 \pi$ which we shall take to be the interval $[-\pi,\pi)$. Then
\begin{equation}\label{phi1}
    \phi = {\rm Tr}(\hat{\phi}\,\hat{\Delta}(R,\phi)){ \longleftrightarrow}
    \hat{\phi} = \int_0^\infty{\rm d}R R
\int_{-\pi}^\pi\frac{{\rm d}\phi}{2 \pi}\,\phi\,
\hat{\Delta}(R,\phi)\,,
\end{equation}
where we now consider $\hat{\Delta}$ as a function of plane polar
coordinates $R$ and $\phi$ so that the matrix elements of
$\hat{\phi}$ with respect to the energy eigen states $|h_n\rangle$,
$(n \geq 0)$ of the harmonic oscillator, with energies $(n +
1/2)\hbar\omega$, are given by
\begin{equation}\label{phimn1}
    \langle h_m|\hat{\phi}| h_n\rangle =
    \int_0^\infty {\rm d}R\, R \int_{-\pi}^{\pi}\frac{{\rm d}\phi}{2
    \pi}\,\phi \,\langle h_m|\,\hat{\Delta}(R,\phi)\,| h_n\rangle\,,
\end{equation}
where \cite{dhsrims1}
\begin{equation}\label{deltamn}
    \langle h_m|\,\hat{\Delta}(R,\phi)\,| h_n\rangle =
    2 (-1)^n \,{\rm i}^{|m-n|}\, 2^{\frac{|m-n|}{2}}\sqrt{\frac{n_\ell !}{n_g
    !}}\,\,{\rm e}^{{\rm i}(n-m)\phi}R^{|m-n|}\, {\rm
    e}^{-R^2}L^{|m-n|}_{n_\ell}(2 R^2)\,.
\end{equation}
Here $n_\ell \,(n_g)$ is the lessor (greater) of the pair $(m,n)$, and
$L_a^b$ is the Laguerre polynomial \cite{gradsht}.  When
(\ref{deltamn}) is used in (\ref{phimn1}) one finds after some
manipulation
\cite{dhsbook, dhsrims1} that
\begin{equation}\label{phimn2}
    \langle h_m|\,\hat{\phi}| h_n\rangle =
    (1-\delta_{m,n})\, \frac{{\rm i}^{n-m+1}}{m-n}\, g_{m,n}
\end{equation}
where $g_{m,n}$ is the symmetric matrix
\begin{equation}\label{gmn1}
    g_{m,n} = 2^{-\frac{|m-n|}{2}}\,
    \frac{\Gamma \left( \frac{n_\ell}{2}+ s_\ell\right)}
    {\Gamma \left( \frac{n_g}{2}+ s_\ell\right)}\sqrt{\frac{n_g !}{n_\ell
    !}}
\end{equation}
with
\begin{equation}\label{gmn2}
   s_{\ell} = \left\{ \begin{array}
   {r@{\quad n_\ell}l}
   1/2& \,\,\, \mbox{even}\\
    1\,\,\, & \,\,\,\mbox{odd}
    \end{array}\right.
\end{equation}

Alternatives to $\hat{\phi}$ have been suggested to represent the phase of an harmonic oscillator, or in some sense photons, for instance by a non-projective positive operator valued measure, or POVM \cite{pegg,pellon1,pellon2,hel,DKPY,KJ}. Using a POVM to represent a quantum system may be thought of as allowing for an element of imperfection in the measurement.
\subsection{Forcing of angle functions by white noise}\label{forcing}
Here, let us consider those operators, $\hat{\Phi}$, whose Weyl quantizations are functions only of angle, $\Phi(\phi)$, in the phase plane, thus
\begin{equation*}
  \hat{\Phi}{ \longleftrightarrow}\Phi(\phi)
\end{equation*}
 Then, let us ask for the expectation of such a $\hat{\Phi}$ when the oscillator is initially in the ground state, $h_0$, and subsequently driven by stochastic Hamiltonian (\ref{randomham}) with (\ref{noise}), namely
\begin{equation}\label{rhoPhione}
  \overline{{\rm Tr}(\hat{\rho}(t)\hat{\Phi})}=
   \int \frac{{\rm d}p\,{\rm d}q}{h} \Phi(p,q)\int {\rm d}p'{\rm d}q'\,
   W(p,q,t|p',q',0)\, \rho_w (p',q',0)\,,
\end{equation}
where
\[\rho_w(p,q,0) = \left(|h_0 \rangle\langle h_0|\right)(p,q) \] is the Weyl transform of the initial state.
The analysis of equation (\ref{rhoPhione}) is similar to that used on equation (\ref{average}). The result is
\begin{eqnarray}\label{rhoPhitwo}
  \overline{{\rm Tr}(\hat{\rho}(t)\hat{\Phi})} &=& \frac{1}{\sqrt{{\rm det}\mathbf A}}
 \int_{- \pi}^\pi \frac{{\rm d}\phi}{2 \pi}\Phi(\phi)
 \int_0^\infty {\rm d}R \, R \,
  {\rm exp}\left[-\frac{R^2}{4 {\rm det}{\mathbf A}}\Big(1+N \omega t - N \sin\omega t \cos(2 \phi - \omega t)\Big)\right] \nonumber\\ \nonumber\\
    &=& \int_{- \pi}^\pi\frac{{\rm d}\phi}{2 \pi}\Phi(\phi)
    \frac{\sqrt{(1+N \omega t)^2 -N^2 \sin ^2\omega t }}{\left[(1+N \omega t) - N \sin\omega t \cos(2 \phi - \omega t)\right]} \equiv \int_{- \pi}^\pi\frac{{\rm d}\phi}{2 \pi}\Phi(\phi)P(\phi,\omega t)\,,
\end{eqnarray}
where we have expanded
\begin{equation}\label{detA}
 4\, {\rm det}\mathbf A = (1+N \omega t)^2 - N^2 \sin^2 \omega t \,.
\end{equation}
Figure 1 shows a computed plot for the choice $\Phi(\phi) = \phi$.

 The factor $P(\phi,\omega t)/(2 \pi)$ in (\ref{rhoPhitwo}) is a bone fide probability density. To see this, first note that, for all $t\geq 0$, it is positive. Secondly, when  $\sin \omega t$ vanishes, and also in the limit $t\rightarrow \infty$, it reverts to $1/2 \pi$, and so corresponds to a random distribution of phase and is manifestly normalized with respect to the interval $[-\pi,\pi)$. When $\sin \omega t$ does not vanish we can, after some algebra, re-express the integral (\ref{rhoPhitwo}) (with $\Phi(\phi)=1$) in terms of the variable $z = {\rm exp}(2 {\rm i}\phi)$ as
\begin{equation}\label{rhoPhithree}
  \int_{- \pi}^\pi\frac{{\rm d}\phi}{2 \pi}P(\phi,\omega t) =-{\rm sign}(\sin \omega t)\,{\rm exp}({\rm i}\omega t) \int_C \frac{{\rm d}z}{2 \pi {\rm i}}
  \frac{\sqrt{\lambda^2 -1}}{(z-z^+)(z-z^-)}
\end{equation}
where the curve $C$ is taken {\em twice} anti-clockwise around the unit circle in the $z$-plane, and
\[z^\pm = {\rm exp}({\rm i}\omega t)(\lambda \pm \sqrt{\lambda^2 -1}) \quad \mbox{with}\quad \lambda=\frac{1 + N \omega t}{N \sin \omega t}\,. \]
Finally, the integral in (\ref{rhoPhithree}) can be seen to be unity, for when ${\rm sign}(\sin \omega t)=+1$, $z^+$ lies within the unit circle and $z^-$ lies without, but when ${\rm sign}(\sin \omega t)=-1$ these roles are exchanged.
\subsection{Forcing by noise at long times}
Equation (\ref{rhoPhitwo}), an exact result when the initial state is $|h_0\rangle$, indicates that the random process $\lambda(t)$ randomizes the phase phase distribution after long times. One might reasonably expect this to occur whatever the initial state, $\rho_w(p,q,0)$. To verify this, from equation (\ref{rhoPhione}) it is clear that we need the limiting behavior of $W(p,q,t|p',q',0)$ for large $t$.

The general expression for $W(|)$ is given by (\ref{osc}). For long times we can make the approximation
\[\int_0^t {\rm d}s \left( \frac{a}{m \omega}\sin \omega s + b \cos\omega s\right)^2
\simeq \frac{t}{2}\left[\left(\frac{a}{m \omega} \right)^2 + b^2 \right]\,.\]
In this limit, the integrals over $a$ and $b$ in (\ref{osc}) reduce to the product of two simple Gaussian integrals. Expressed in terms of the dimensionless variables $x$ and $y$ the long-time limiting behavior of $W$ is
\begin{equation*}\label{Wlongt}
  W(x,y,t|x',y',0) \simeq \frac{1}{\pi \hbar N \omega t}\,{\rm exp} \left[-\frac{1}{N \omega t}(x - x_t)^2 +(y - y_t)^2 \right]\,,
\end{equation*}
where $x_t$ and $y_t$ are the oscillator's free motion with initial values $x'$ and $y'$ at time zero. In polar coordinates we can write
\[x+{\rm i}\,y = R\,{\rm exp}{\rm i} \phi \quad \mbox{and}\quad
x_t+{\rm i}\,y_t = R' \, {\rm exp}{\rm i} (\phi ' + \omega t)\,, \]
so that
\[(x - x_t)^2 +(y - y_t)^2 = R^2 + R'^2 -2RR' \cos(\phi - \phi' - \omega t)\,. \]
Then using this in (\ref{rhoPhione}) gives (for long times) for any operator
\[\hat{O}{ \longleftrightarrow}O(p,q)\,, \]
\begin{eqnarray}\label{rhoPhifour}
  \overline{{\rm Tr}(\hat{\rho}(t)\hat{O})} & \longrightarrow & \int_{-\pi}^\pi \frac{{\rm d}\phi}{2 \pi}
\int_0^\infty {\rm d}R R \,\, O(R,\phi) \\ & &
  \hspace{-3cm} \times \frac{1}{N \omega t} \int_0^\infty {\rm d}R' R' \int_{-\pi}^\pi \frac{{\rm d} \phi'}{\pi}
{\rm exp}\left[- \frac{1}{N \omega t}\left( R^2 + R'^2 -2RR' \cos(\phi - \phi' - \omega t)\right) \right]
\rho_w(R',\phi',0)\nonumber\,.
\end{eqnarray}
In particular, for those operators $\hat{\Phi}$ whose Weyl equivalents $\Phi(\phi)$ are functions only of $\phi$ we can write (changing variables to $x = R/\sqrt{N \omega t}$)
\begin{eqnarray*}
  \overline{{\rm Tr}(\hat{\rho}(t)\hat{\Phi})} & \longrightarrow & \int_{-\pi}^\pi \frac{{\rm d}\phi}{2 \pi} \Phi(\phi)\int_0^\infty {\rm d}x\, x  \\ & &
  \hspace{-2.5cm} \times \int_0^\infty {\rm d}R' R' \int_{-\pi}^\pi
\frac{{\rm d}\phi'}{\pi}
{\rm exp}\left[ -\left( x^2 + \frac{R'^2}{N \omega t} -\frac{x R'}{\sqrt{N \omega t}}\cos(\phi -\phi' -\omega t) \right)\right] \rho_w(R',\phi',0) \\  \\
    &\longrightarrow& \int_{-\pi}^\pi \frac{{\rm d}\phi}{2 \pi} \Phi(\phi)
 \int_0^\infty {\rm d}x\, x \int_0^\infty {\rm d}R' R' \int_{-\pi}^\pi\frac{{\rm d}\phi'}{\pi}
{\rm exp}\left( - x^2\right) \rho_w(R',\phi',0)\\ \\
   &=&\int_{-\pi}^\pi \frac{{\rm d}\phi}{2 \pi} \Phi(\phi)
   \int_0^\infty {\rm d}R' R' \int_{-\pi}^\pi \frac{{\rm d}\phi'}{2 \pi}
   \, \rho_w(R',\phi',0)= \int_{-\pi}^\pi \frac{{\rm d}\phi}{2 \pi} \Phi(\phi)\, {\rm Tr}(\hat{\rho}(0))\\ \\
   &=& \int_{-\pi}^\pi \frac{{\rm d}\phi}{2 \pi} \Phi(\phi)\,.
\end{eqnarray*}
Thus for these operators the phase distribution is randomized in time by the stochastic force.

Similarly, we may ask for the long-time effect on operators whose Weyl equivalents are functions only of $R$, namely
those $\Omega(R)$ such that
\[\hat{\Omega} { \longleftrightarrow} \Omega(R)\,.\]
In this case, the same sort of formal manipulations give
\begin{eqnarray}\label{rhoPhifive}
  \overline{{\rm Tr}(\hat{\rho}(t)\hat{\Omega})} & \longrightarrow & \int_{-\pi}^\pi \frac{{\rm d}\phi}{2 \pi} \int_0^\infty {\rm d}x\, x \, \Omega(x\sqrt{N \omega t})\nonumber \\ & &
  \hspace{-2.5cm} \times \int_0^\infty {\rm d}R' R' \int_{-\pi}^\pi
\frac{{\rm d}\phi'}{\pi}
{\rm exp}\left[ -\left( x^2 + \frac{R'^2}{N \omega t} -\frac{x R'}{\sqrt{N \omega t}}\cos(\phi -\phi' -\omega t) \right)\right] \rho_w(R',\phi',0)\nonumber \\
    &\longrightarrow&
    \int_{-\pi}^\pi \frac{{\rm d}\phi}{2 \pi} \int_0^\infty {\rm d}                                         x x \,\Omega(x \sqrt{N \omega t})\,2 \, {\rm exp}(-x^2)\, {\rm Tr}(\rho(0))\nonumber\\
   &=& \int_0^\infty {\rm d}x \, {\rm exp}(-x)\, \Omega(\sqrt{x\,N \omega t })\,.
\end{eqnarray}
For instance if we choose $\hat{\Omega} = {\rm exp}(-\beta \hat{H})$, where $\beta$ is a parameter and $\hat{H}= (\hat{n} + 1/2)\hbar \omega$ is the Hamiltonian for a free oscillator, then, from equation (\ref{deltamn}) and reference to standard tables (eg \cite{gradsht}),
\begin{eqnarray*}
{\rm exp}(-\beta \hat{H}) &{ \longleftrightarrow}& {\rm Tr} \left({\rm exp}(-\beta \hat{H})
\hat{\Delta}(R,\phi)\right)\\
                          &=& 2 \sum_{m=0}^\infty {\rm exp}\big(-(m+1/2)\hbar \omega \beta \big)
(-)^m {\rm exp}(-R^2)L_m(2 R^2)\\
                          &=& \frac{1}{\cosh \left(\frac{\hbar \omega \beta}{2}\right)}
                          {\exp}\left[-R^2 \tanh\left(\frac{\hbar \omega \beta}{2}\right)\right]\,.
\end{eqnarray*}
Thus, in this instance we have, using (\ref{rhoPhifive}) and considering the limiting behavior for large $t$,
\begin{equation}\label{beta}
 \overline{{\rm Tr}\left(\hat{\rho}(t){\rm exp}(-\beta \hat{H})\right)}\longrightarrow \left[\cosh\left(\frac{\hbar \omega \beta}{2}\right)+ N\omega t \sinh\left(\frac{\hbar \omega \beta}{2}\right)\right]^{-1}\,.
\end{equation}

Although the Weyl equivalent to $\hat{\phi}$ is the pure phase $\phi$, the Weyl equivalents of powers and functions of $\hat{\phi}$ are often not functions of $\phi$ only. Indeed, even the Weyl equivalent of the square $\hat{\phi}^2$ is a mixed function of $\phi$ and $R$. One can show (\cite{sdh92}) using (\ref{phimn2}) and choosing the range $[-\pi,\pi)$, that
\begin{equation}\label{sigmathree}
  \langle h_m|\hat{\phi}^2 | h_m\rangle =
  \sum_{n=0}^\infty |\langle h_m|\hat{\phi}|h_n \rangle|^2 =
  \sum_{n=1}^m \frac{1}{n^2}\, (g_{m,m-n})^2 +
  \sum_{n=1}^\infty \frac{1}{n^2}\,(g_{m+n,m})^2\,,
\end{equation}
where the first sum on the right-hand side only contributes when $n\geq 1$.
It is, however, a property of $g_{m,n}$, equation (\ref{gmn1}), that as both $m$ and $n$ increase to values that are large with respect to their difference (the correspondence limit) then $g_{m,n} \rightarrow 1$.  Thus as $m$ increases in (\ref{sigmathree}) the standard deviation $\langle h_m|\hat{\phi}^2 | h_m\rangle$ approaches $\pi^2/3$, a value associated with a {\em random} distribution of phase (\cite{sdh92}).

We may ask what effect the random force $\lambda(t)$ might have on $\hat{\phi}^2$. In particular, if its Weyl transform, in polar coordinates, is
\begin{equation}\label{phisq}
( \hat{\phi}^2)(R,\phi)=
{\rm Tr}\left(\hat{\Delta}(R,\phi)\,\hat{\phi}^2\right)\,,
\end{equation}
then, adapting the development leading to (\ref{rhoPhifive}), we have (for long times)
\begin{equation}\label{phisqlim}
  \overline{{\rm Tr}(\hat{\rho}(t)\hat{\phi}^2)}  \longrightarrow
  2\int_0^\infty {\rm d}x\, x \,{\rm exp}(-x^2)\int_{-\pi}^\pi\frac{{\rm d}\phi}{2\pi}\,
  \big( \hat{\phi}^2\big)(x \sqrt{N \omega t},\phi)\,.
\end{equation}
Now, from (\ref{deltamn}) and (\ref{phisq}),
\begin{eqnarray}
\nonumber \int_{-\pi}^\pi \frac{{\rm d}\phi}{2\pi} ( \hat{\phi}^2 )(R,\phi) &=& \sum_{m=0}^\infty \left( \langle h_m|\hat{\phi}^2 | h_m\rangle-\frac{\pi^2}{3}+\frac{\pi^2}{3}\right)\, 2 (-)^m {\rm exp}(-R^2)L_m (2 R^2) \\ \nonumber &=&\frac{\pi^2}{3}+\sum_{m=0}^\infty \left( \langle h_m|\hat{\phi}^2 | h_m\rangle-\frac{\pi^2}{3}\right)\, 2 (-)^m {\rm exp}(-R^2)L_m (2 R^2)\,,
\end{eqnarray}
where I have used (in the limit $z\rightarrow -1$) the equation \cite{gradsht}
\[(1-z)^{-1}{\rm exp}\left( \frac{xz}{z-1} \right) =\sum_{m=0}^\infty L_m(x) z^m\,. \]
Using this in equation (\ref{phisqlim}) shows that for long times
\begin{equation}\label{phitwolim}
  \overline{{\rm Tr}(\hat{\rho}(t)\hat{\phi}^2)}  \longrightarrow \frac{\pi^2}{3}\,.
\end{equation}
This is characteristic of a random distribution of $\phi$ over the interval $[-\pi,\pi)$, namely
\[\int_{-\pi}^\pi\frac{{\rm d}\phi}{2\pi} \phi^2 = \frac{\pi^2}{3}\,. \]
\section{Time dependent frequency}\label{tdepfreq}
\subsection{Adiabatic frequency dependence}
We apply the `method of averaging' \cite{dz} to the oscillator, equations (\ref{motion}) when there is no external force but with a time-dependent frequency,
\begin{equation}\label{freq}
  \omega^2(t) = \omega_0^2 \big(1+ \epsilon (t)\big)\,,
\end{equation}
where $\omega_0$ is a constant base frequency and we will ultimately assume that $\epsilon(t)$ is very small compared to unity. Then defining variables $(x,y) = (p/(\hbar \alpha),\alpha q)$, where $\alpha ^2 \equiv m \omega_0  /\hbar \,, $
we may opt to express the (unknown) solution to the equations of motion in the form
\begin{equation}\label{rad}
  x(t) + {\rm i}\, y(t) = r(t)\, {\rm e}^{{\rm i}(\omega_0 t +\theta(t)) }\,.
\end{equation}
Then
\begin{equation}\label{one}
  \dot x(t) +{\rm i}\, \dot y(t) =\left[\dot r(t) + {\rm i}\,r(t)\big(\omega_0 +
\dot \theta (t))\right]{\rm e}^{{\rm i}(\omega_0 t +\theta(t))}\,.
\end{equation}
This equation is a result of the choice (\ref{rad}). But the motion must be governed by the equations
\[ p(t)=m \dot q (t)\quad \mbox{and}\quad \dot p (t) = -m \omega^2(t) q(t)\,.\]
These require that
\[\dot x(t) +{\rm i}\dot y(t) = \frac{\dot p(t)}{\hbar \alpha}+{\rm i}\,\alpha\, \dot q(t) =
-\frac{m }{\hbar \alpha}\omega^2(t)q(t)  +{\rm i}\frac{\alpha}{m}p(t)\,,\]
which can rearranged to give
\begin{eqnarray}\label{two}
  \dot x(t) +{\rm i}\dot y(t) &=& {\rm i}\,\omega_0 \big(x(t)+{\rm i}y(t)+
{\rm i}\,\epsilon(t) y(t) \big)\nonumber  \\
   &=& {\rm i}\,\omega_0\, r(t){\rm e}^{{\rm i}(\omega_0 t +\theta(t))}
   -\omega_0 \epsilon(t) y(t)
\end{eqnarray}
Equating the right-hand sides of (\ref{one}) with (\ref{rad}), and (\ref{two}), simplifying, and then separating the real and imaginary parts gives
\begin{eqnarray}\label{rtwodtwo}
  \dot{r}(t) &=& - \frac{1}{2}\,\omega_0\, r(t)\, \epsilon(t)
  \sin\big( 2(\omega_0 t + \theta(t)\big) \nonumber  \\ \\
  \dot{\theta}(t) &=& \omega_0\, \epsilon(t) \sin ^2 (\omega_0 t + \theta(t)) \nonumber\,.
\end{eqnarray}
 For the adiabatic limit we assume that base period, $2 \pi/ \omega_0$, is very much smaller than the time over which $\epsilon(t)$ varies significantly. In that case, denoting by $\theta_a(t)$ and $r_a(t)$ a `local' or adiabatic time average of $\theta$ and $r$ over a time long compared to the base period $2 \pi/ \omega_0$ but short compared to $\epsilon(t)/\dot{\epsilon}(t)$ we have
\begin{equation*}
  \dot{r}_a(t) = 0 \quad \mbox{and} \quad \dot{\theta}_a(t) = \frac{1}{2}\,\omega_0 \epsilon(t)\,,
\end{equation*}
so that
\begin{eqnarray}
  r_a(t) &=& r(0)\,\,\, \mbox{and} \\
  \theta_a(t) & = & \theta(0) + e(t),\,\,\, \mbox{where}\\
  e(t) & \equiv & \frac{\omega_0}{2}\int_0^t {\rm d}s \, \epsilon(s)\,.
\end{eqnarray}
To this level of approximation the simple time dependent phase shift $e(t)$ will not affect transition probabilities.
\subsection{Parametric oscillator}
Suppose the frequency varies at approximately twice the fundamental rate $\omega_0$, so that
\begin{equation}\label{Paroscone}
  \omega^2(t) = \omega_0^2 [1 + \overline{e} \cos\big(2(\omega_0 + f)t \big)]\,,
\end{equation}
and that $\overline{e}$ and $f/\omega_0$ are small compared to unity. Then, assuming an approximate solution in the form of equation (\ref{rad}), and defining the parameter $u \equiv \overline{e} \omega_0/4$, equations (\ref{rtwodtwo}) take the form
\begin{eqnarray*}
  \dot{r}(t) &=&-2 u \, r(t)\cos((2\omega_0 +f)t )\sin(2(\omega_0 t -f t)) \\
  \dot{\theta}(t) &=& 4 u  \cos((2\omega_0 +f)t) \sin^2(\omega_0 t +\theta(t))\,.
\end{eqnarray*}
Expanding the right-hand sides of these equations using basic trigonometric identities, and retaining only the slowly varying terms gives
\begin{eqnarray*}
  \dot{r}(t) & \simeq & - u \, r(t) \sin(2 \theta(t) -f t) \\
  \dot{\theta}(t) & \simeq & - u \, \cos(2 \theta(t) -f t) \,.
\end{eqnarray*}
This model is discussed in reference \cite{Land}, where it is shown that growth occurs in the classical motion for values of $f$ such that $-2 u < f < 2 u$. For simplicity we choose to pursue the case $f=0$. Then the equations to solve are
\begin{eqnarray}\label{rtheta}
  \dot{r} & \simeq & - u \, r(t) \sin(2 \theta(t)) \\
  \dot{\theta}(t) &\simeq& - u \cos(2 \theta(t))\,.
\end{eqnarray}
The second of these can be solved by direct integration:
\begin{equation}\label{theta}
  \tan\left(\theta(t) +\frac{\pi}{4}\right )= {\rm exp}(-2 u t)\tan\left(\phi_0 +\frac{\pi}{4}\right )
\end{equation}
where $\theta(0) = \phi_0$ is the initial angle in the phase plane for the oscillator's motion.
Consistent with this we assume the following solution for $x(t)+ {\rm i}\,y(t)$:
\begin{equation}\label{rsoln}
  x(t)+ {\rm i}\,y(t)= R_0 \,{\rm exp}\left({\rm i}(\omega_0 t - \frac{\pi}{4}) \right)
  \left[ {\rm exp}({u t}) \cos\left(\phi_0 + \frac{\pi}{4}\right)+{\rm i}\, {\rm exp}(-{u t}) \sin \left(\phi_0 + \frac{\pi}{4}\right) \right]\,,
\end{equation}
where $R_0$ is the initial radius in polar coordinates. Then
\begin{equation}\label{radius}
  r(t)=R_0 \sqrt{{\rm exp}(2 u t) \cos^2 \left(\phi_0 +\frac{\pi}{4}\right)+
  {\rm exp}(-2 u t) \sin^2 \left(\phi_0 +\frac{\pi}{4} \right)}\,.
\end{equation}
Differentiating this with respect to time and using equation (\ref{theta}) shows that this is a solution to (\ref{rtheta}).

One can of course calculate transition probabilities for this model as in subsection \ref{transprob}. In particular, by equation (\ref{trans}), the probability that the oscillator stays in the ground state after time $t$ is
\begin{equation}\label{transpara}
  |\langle h_0 |\hat{U}_t | h_0 \rangle|^2=\int \frac{{\rm d}p\,{\rm d}q}{h}\int {\rm d}p'{\rm d}q'\, \big(|h_0\rangle\langle h_0|\big)(p,q)P_w (p,q,t|p',q',0)\big(|h_0\rangle\langle h_0|\big)(p',q')\,,
\end{equation}
where
\[{\rm d}p\,{\rm d}q = \hbar\, {\rm d}x \,{\rm d}y=\hbar\, R\, {\rm d} R\, {\rm d} \phi\,, \]
and, recalling equation (\ref{ground}),
\[ (|h_0 \rangle\langle h_0|)(p,q)=2 \,{\rm exp}\left(- (x^2 +y^2)\right)=2\, {\rm exp}(-R^2)\,. \]
Now
\[P_w (p,q,t|p',q',0)=\delta(p-p(t))\delta(q-q(t))=\frac{1}{\hbar R}\delta(R-r(t)\delta(\phi-\theta(t))\,. \]
where, $(r(t),\theta(t))$ are the solutions for the parametric oscillator in polar coordinates with initial conditions $(R',\phi')$.
The Weyl transform of $|h_0 \rangle\langle h_0|$ depends only on the radial coordinate and for this motion, using (\ref{radius}),
\[r^2(t)=R'^2 \left[ {\rm exp}(2 u t) \cos^2 \left(\phi' + \frac{\pi}{4}\right)
+ {\rm exp}(-2 u t) \sin^2 \left(\phi' + \frac{\pi}{4} \right) \right]\,. \]
With this information (\ref{transpara}) simplifies to
\begin{eqnarray*}
  |\langle h_0 |\hat{U}_t | h_0 \rangle|^2 &=& 4\int_0^\infty {\rm d}R' \int_{-\pi}^\pi
  \frac{{\rm d} \phi'}{2 \pi} {\rm exp}\left[-R'^2 \left( {\rm exp}(2 u t) \cos^2 \left(\phi' + \frac{\pi}{4}\right) + {\rm exp}(-2 u t) \sin^2 \left(\phi' + \frac{\pi}{4} \right) \right)\right] \\ \\
   &=& \int_{-\pi}^\pi \frac{{\rm d}\phi'}{\pi}\frac{1}{1+ {\rm exp}(2 u t) \cos^2 \left(\phi' + \frac{\pi}{4}\right) + {\rm exp}(-2 u t) \sin^2 \left(\phi' + \frac{\pi}{4} \right) }\\ \\
   &=& \int_{-\pi}^\pi \frac{{\rm d} \phi'}{ \pi}\frac{1}{\sinh(2 u t)\left( \cos(2 \phi)+\coth(u t) \right)}
\end{eqnarray*}
To evaluate the last integral here we can, as before in subsection \ref{forcing}, let
$z={\rm exp}(2 {\rm i}\phi)$ and integrate with respect to $z$ {\em twice} around the unit circle in the complex $z$-plane, so giving (with some manipulation)
\begin{equation*}
  |\langle h_0 |\hat{U}_t | h_0 \rangle|^2 = \frac{2}{\sinh(2 u t)}
  \int_C \frac{{\rm d}z}{2 \pi {\rm i}}\frac{1}{(z-z+)(z-z^-)}\,,
\end{equation*}
where $z^+ = - \tanh(u t/2)$ and $z^- = -\coth(u t/2)$. Examination shows that $z^+$ lies within the unit circle and $z^-$ without, so that, by standard means (and after a little simplification)
\begin{equation}\label{paraprob}
  |\langle h_0 |\hat{U}_t | h_0 \rangle|^2 = \frac{1}{\cosh(u t)}\,.
\end{equation}
In this case, then, the probability that the oscillator stays in the ground state decays exponentially at long times.
\section{Discussion}\label{disc}
Ultimately, quantum mechanics calculates expectations and probabilities through the density matrix $\hat{\rho}(t)$ which always involves {\em products} of wave functions.  The Wigner function $\rho_w(p,q;t)$ was originally devised \cite{wigner} as a scheme for using $\hat{\rho}$ to extract from the density matrix quantum corrections to classical physics. In precise terms, the Wigner function is the Weyl transform of $\hat{\rho}/h$, that is to say $\rho_w  \longleftrightarrow \hat{\rho}(t)/h$.
For systems govererned by generally time-dependent {\em quadratic} Hamiltonians the time-development of $\rho_w(p,q;t)$ is, at its core, purely classical, with the fully quantum result to be be found through equations (\ref{opevo}) and (\ref{classicalP}).

Here, we have drawn upon this quantum formulation to calculate exactly several expectations and probabilities for the harmonic oscillator when acted upon by an external white noise force. To do this one must calculate the ensemble average of $\hat{\rho}$ and not of the wave functions themselves.  As one might expect, when the oscillator is initially in its ground state, over time the random force pumps it up to higher energy levels, statistically speaking, according to equation (\ref{quadfour}).

The ultimate randomizing effect of the stochastic force on the Weyl quantization of the oscillator's angle $\hat{\phi} \longleftrightarrow \phi$ in the phase plane, and indeed of all functions $\hat{\Phi} \longleftrightarrow \Phi(\phi)$, is reasonable: over time, whatever its initial state $\hat{\rho}(0)$, the distribution of angle is `smeared' to uniformity over the interval $[-\pi,\pi)$.  By the same token one might expect that the same force would, in time, pump up the oscillator's energy. The quantum statement of this is given by (\ref{beta}), which shows, for any initial state, that large times
\[{\rm Tr}\big(\hat{\rho}(t) {\rm exp}(-\beta \hat{H}) \big)\sim \frac{1}{t}\,, \]
or, expanding equation (\ref{beta}) in powers of $\beta$ gives to first order, that
\[{\rm Tr}\big(\hat{\rho}(t) \hat{H}\big)\rightarrow \frac{\hbar \omega}{2}N \omega t \]
where the strength of the random force is characterized by the dimensionless parameter $\sqrt{N}$, equation (\ref{defs}).

We also considered, in section \ref{tdepfreq}, an oscillator subjected not to any external force, but to a small time-dependent perturbation in its fundamental frequency. When the dependence is slow compared to the oscillator's natural frequency---the adiabatic limit---the resultant effect on probabilities is small. But when, as for a parametric oscillator, this sinusoidal time-dependence has about twice the oscillator's natural frequency, though the perturbation may be weak, the effects are profound.  Equation (\ref{paraprob}) shows for instance that the probability that the oscillator stays in its ground state decays, for long times, exponentially.
\subsection*{Acknowledgement}
The author thanks Dr. D. A. Dubin for enlightening conversations about the Weyl quantized phase.
\newpage
\section{Figures}
\begin{figure}[h]
\begin{center}
\includegraphics{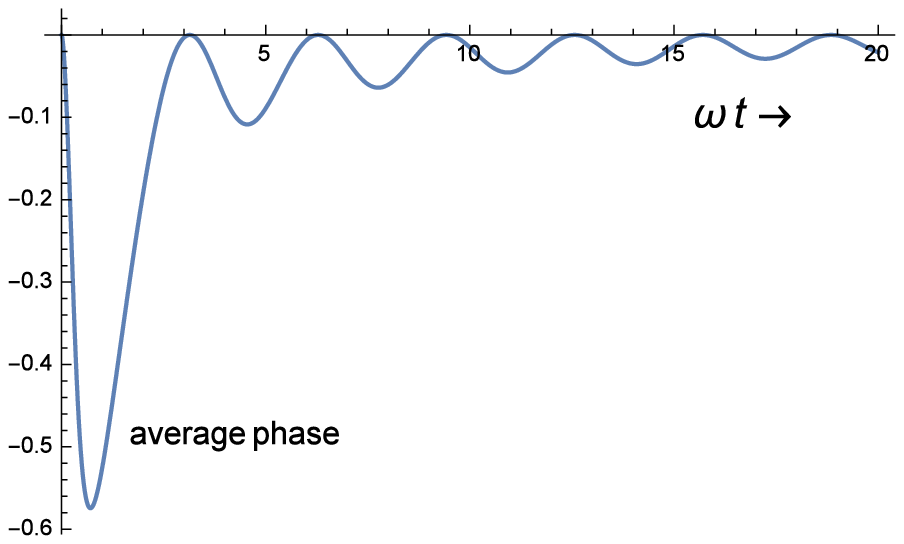}
\caption{\label{avephi}  The expectation of phase (equation (\ref{rhoPhitwo}) with $\Phi(\phi)=\phi$) with $N=1$, and $\omega t$ ranging from $0$ to $20$.}
\end{center}
\end{figure}

\end{document}